\documentclass[english,aps,prd,%twocolumn,
groupedaddress,amssymb,
showpacs,nofootinbib]{revtex4}

\usepackage{times}
\usepackage{amsmath}
\usepackage{graphicx}

\newcommand{\beq}{\begin{eqnarray}}
\newcommand{\eeq}{\end{eqnarray}}

\makeatletter
\numberwithin{equation}{section}
\numberwithin{figure}{section}

\makeatother

\usepackage{babel}

\begin{document}
\setlength{\unitlength}{1mm}

\title{Nontrivial Thermodynamics in  't Hooft's Large-$N$ Limit}

\author{Axel \surname{Cort\'es Cubero}}

\email{acortes@sissa.it}

\affiliation{ Scuola Internazionale Superiore di Studi Avanzati (SISSA), and INFN, Sezione di Trieste; via Bonomea 265, 34136 Trieste, Italy
}

\begin{abstract}

We study the finite volume/temperature correlation functions of the (1+1)-dimensional ${\rm SU}(N)$ principal chiral sigma model in the planar limit. The exact S-matrix of the sigma model is known to simplify drastically at large $N$, and this leads to trivial thermodynamic Bethe ansatz (TBA) equations. The partition function, if derived using the TBA, can be shown to be that of free particles. We show that the correlation functions and expectation values of operators at finite volume/temperature are not those of the free theory, and that the TBA does not give enough information to calculate them. 
Our analysis is done using the Leclair-Mussardo formula for finite-volume correlators, and knowledge of the exact infinite-volume form factors. We present analytical results for the one-point function of the energy-momentum tensor, and the two-point function of the renormalized field operator. The results for the energy-momentum tensor can be used to define a nontrivial partition function.

\end{abstract}


\maketitle

\section{Introduction}

One of the main goals of statistical physics is to calculate expectation values of observables in a system at finite temperature.
 The partition function usually contains enough information to find some of these  expectation values, which can be computed by taking different partial derivatives of it. 
The most common tool used to derive the finite-temperature partition function of a two-dimensional integrable quantum field theory is the thermodynamic Bethe ansatz (TBA). In this paper we propose that this tool does not work in general for a matrix-valued quantum field theory. The expectation values of operators include additional information not contained in the TBA partition function. 

To demonstrate this proposal, we examine the (1+1)-dimensional principal chiral sigma model (PCSM).   
The PCSM has the action
\beq
S_{\rm PCSM}=\int d^2 x\,\frac{1}{2g_0^2}{\rm Tr}\partial_\mu U^\dag(x)\partial^\mu U(x),\label{pcsmaction}
\eeq
where $U(x)\in {\rm SU}(N)$. This model has been shown to be integrable, and its exact S-matrix is known \cite{wiegmann}. The action (\ref{pcsmaction}) has an ${\rm SU}(N)\times {\rm SU}(N)$ global symmetry given by $U(x)\to V_L U(x) V_{R}$, with $V_{L,R}\in {\rm SU}(N)$. The PCSM is asymptotically free and has a mass gap, which we call $m$. In our analysis, we simply assume the existence of a mass gap. A mechanism explaining how this mass is dynamically generated was proposed in \cite{unsal}.

We are interested particularly in 't Hooft's large-$N$ limit of the PCSM.  In this limit, the S-matrix greatly simplifies. We later show that this means that the TBA partition function at large $N$ is that of a free theory. However, we show that the expectation values of operators are not trivial. 

The inefficiency of this partition function is due to the matrix structure of the theory. It has been shown that the form factors of operators are not trivial at large $N$ \cite{renormalizedfield}\cite{multiparticle}\cite{correlation}. As we will see, the computation of thermal expectation values can be done by summing over form factors. Since these form factors are not trivial (despite the trivial S-matrix), the thermal expectation values are not trivial either (despite the trivial TBA).

In the rest of this paper we will use interchangeably the terms finite volume, and finite temperature. This is because in 1+1 dimensions, these two are equivalent up to a Wick rotation. 

In the next section we review some of the exact results that are known for the PCSM. In Section III we show how the form factors have been used before to calculate infinite-volume correlation functions of operators, which agree with the asymptotic freedom of the PCSM. We discuss the results of Ref. \cite{asymptoticfreedom}, where the infinite-volume two-point function of the renormalized-field operator was computed.

 In Section IV, we discuss the application of the TBA to the PCSM. We find that the TBA yields the partition function of a free field at large $N$.
 
In Section V we compute the one-point function (vacuum expectation value) of the trace of the energy-momentum tensor operator. This correlation function is computed using the so-called Leclair-Mussardo (LM)  formula \cite{leclairmussardo}. We observe that this one-point function does not agree with what is expected from the trivial TBA.   We then show how to define a nontrivial partition function from our result for the energy-momentum tensor.

 In Section VI, we  compute two-point function of the renormalized-field operator in a finite volume. This is a finite-volume version of the result of \cite{asymptoticfreedom}.  While the LM formula for one-point functions is generally believed to be accurate, the validity of the LM formula for two-point functions has been disputed \cite{saleur}, \cite{fvform}. We argue that the objections from \cite{saleur} and \cite{fvform} do not affect the PCSM at large N, and that the LM formula might be valid in our case (though we have no proof that this is the correct two-point function).  
The very large and very small volume limits of this two-point function are examined in detail in  Section VII.

\section{Form Factors of the Principal Chiral Sigma Model}

In the following two sections we show a brief review of previous results on exact form factors and correlation functions of the PCSM. 

The main tool that has been used in previous works is the form factor bootstrap program for integrable field theories \cite{bootstrap}. The integrability of the PCSM implies that all scattering events are completely elastic and factorizable into a product of two-particle S-matrices. 

All the qualities of an elementary excitation are specified by stating its rapidity $\theta$, related to its energy and momentum by $E=m\cosh\theta$, $p=m\sinh\theta$, its left and right ${\rm SU}(N)$ color indices $a,b=1,\dots,N$, respectively, and by stating if the excitation is a particle or an antiparticle. We can write, for example, a one-particle incoming state and a one-antiparticle incoming state as
\beq
\vert P,\theta,a,b\rangle_{\rm in},\,\,\,\,\,\,\,\,\vert A,\theta,b,a\rangle_{\rm in},\nonumber
\eeq
respectively.

The particle-antiparticle S-matrix, $S(\theta)_{a_1b_1;b_2a_2}^{d_2c_2;c_1d_1}$, defined by 
\beq
\,_{\rm out}\langle A, \theta^\prime_1,d_1,c_1; P,\theta^\prime_2,c_2,d_2\vert A, \theta_1, b_1,a_1; P, \theta_2,a_2,b_2\rangle_{\rm in}=S(\theta)_{a_1b_1;b_2a_2}^{d_2c_2;c_1d_1}\,4\pi \delta(\theta_1-\theta^\prime_1)\,4\pi \delta(\theta_2-\theta^\prime_2),\nonumber
\eeq
 is known to be \cite{wiegmann}
\beq
S(\theta)_{a_1b_1;b_2a_2}^{d_2c_2;c_1d_1}=Q(\theta)\left[\delta_{a_1}^{c_1}\delta_{a_2}^{c_2}-\frac{2\pi {\rm i}}{N(\pi {\rm i}-\theta)}\delta_{a_1a_2}\delta^{c_1c_2}\right]\left[\delta_{b_1}^{d_1}\delta_{b_2}^{d_2}-\frac{2\pi {\rm i}}{N(\pi {\rm i}-\theta)}\delta_{b_1b_2}b^{d_1d_2}\right],\label{sindex}
\eeq
where
\beq
Q(\theta)=\frac{\sinh\left[\frac{(\pi {\rm i}-\theta)}{2}-\frac{\pi {\rm i}}{N}\right]}{\sinh\left[\frac{(\pi {\rm i}-\theta)}{2}+\frac{\pi {\rm i}}{N}\right]}\,\left\{\frac{\Gamma[i(\pi {\rm i}-\theta)/2\pi+1]\Gamma[-{\rm i}(\pi {\rm i}-\theta)/2\pi-{1}/{N}]}{\Gamma[{\rm i}(\pi {\rm i}-\theta)/2\pi+1-1/N]\Gamma[-{\rm i}(\pi {\rm i}-\theta)/2\pi]}\right\}^2,\label{wiegmann}
\eeq
and $\theta=\theta_1-\theta_2$. The particle-particle and antiparticle-antiparticle S-matrices can be found using crossing symmetry. An incoming particle (antiparticle) can be turned into an outgoing antiparticle (particle), by shifting its rapidity by $\theta\to\theta-\pi i$.
For general $N$, there exist $r$-particle bound states, with mass
\beq
m_r=m\frac{\sin\left(\frac{\pi r}{N}\right)}{\sin\left(\frac{\pi}{N}\right)},\,r=1,\dots,N-1.\nonumber
\eeq

For the rest of this paper we will work exclusively in `tHooft's large-$N$ limit. That is, we take $N\to\infty,$ while keeping $m$ fixed.This limit simplifies the problem in many ways. First of all, there are no bound states at large $N$, since the binding energy vanishes. Also the S-matrix is greatly simplified in this limit, as $Q(\theta)=1+\mathcal{O}\left(1/N^2\right)$. As was pointed out in \cite{kazakov}, the thermodynamic Bethe ansatz equations in   `t Hooft's large-$N$ limit are essentially those of a free theory (the authors of this reference later investigate a different large-$N$ limit with nontrivial Bethe equations, which we do not discuss further here). 

At large $N$, two excitations interact nontrivially only if they have color indices contracted with each other. This is easily seen from Eq. (\ref{sindex}). The non symmetric terms in the S-matrix, proportional to $\delta_{a_1 a_2}\delta^{c_1 c_2}$, or $\delta_{b_1 b_2}\delta^{d_1 d_2}$, vanish at large $N$, unless one sums over the colors of one of these delta functions. A particle has a left and a right color index, so it can interact nontrivially with at most two other excitations.

The form factors (matrix elements of local operators) of the renormalized field, $\Phi(x)$, have been found in the large-$N$ limit, in Ref. \cite{renormalizedfield}. This field is defined in terms of the bare field, $U$, by 
\beq
\langle0\vert{\rm Tr}\,\Phi(x)\Phi(0)^\dag\vert0\rangle=Z[g_0(\Lambda),\Lambda]^{-1}\langle0\vert{\rm Tr} \,U(x)U(0)^\dag\vert0\rangle,\nonumber
\eeq
where $Z[g_0(\Lambda),\Lambda]$ is a renormalization constant, $\Lambda$ is the Euclidean momentum cutoff, and $g_0(\Lambda)$ is the coupling constant, which runs such that the mass gap, $m$, is independent of the cutoff.
 We write here form factors with excitations only in the incoming state, as outgoing particles can be obtained using crossing symmetry. Because of the ${\rm SU}(N)\times {\rm SU}(N)$ symmetry of the PCSM, only form factors with $M$ particles and $M-1$ antiparticles are non-vanishing, where $M$ is a positive integer. The  form factors can be parametrized as 
\beq
&\langle0\vert\Phi(0)_{b_0a_0}\vert A,\theta_1,b_1,a_1;\dots;A,\theta_{M-1},b_{M-1},a_{M-1};P,\theta_{M},a_M,b_M;\dots;P,\theta_{2M-1},a_{2M-1},b_{2M-1}\rangle\nonumber\\
&=N^{-M+1/2}\sum_{\sigma,\tau\in S_{M}} F^\Phi_{\sigma\tau}(\theta_1,\dots,\theta_{2M-1})\prod_{j=0}^{M-1}\delta_{a_ja_{\sigma(j)+M}}\delta_{b_jb_{\tau(j)+M}},\label{formfactoransatz}
\eeq
where $\sigma$ is a permutation that takes the set of numbers $0,1,2,\dots,M-1$ to $\sigma(0),\sigma(1),\dots,\sigma(M-1)$, and $\tau$ takes the numbers $0,1,2,\dots,M-1$ to $\tau(0),\tau(1),\dots,\tau(M-1)$, and we sum over all the possible permutations in the set $S_M$. The main result of Ref. \cite{renormalizedfield} is (at large $N$):
\beq
F^{\Phi}_{\sigma\tau}(\theta_1,\dots,\theta_{2M-1})=\left\{
\begin{array}{c}
\frac{(-4\pi)^{M-1}}{\prod_{j=1}^{M-1}[\theta_{j}-\theta_{\sigma(j)+M}+\pi i][\theta_j-\theta_{\tau(j)+M}+\pi i]},\,\,\,\,\sigma(j)\neq\tau(j), {\rm\,for\,all\,}j\\
0,\,\,\,\,{\rm otherwise}\end{array}\right.\,.\label{formfactorsolution}
\eeq
A crucial tool for being able to find these form factors was the simplicity of the S-matrix at large $N$. The scattering of any two incoming excitations in (\ref{formfactoransatz}) is trivial except for the pairs of permutations $\sigma,\,\tau$ where one or both of their color indices are contracted.

We are also interested in the form factors of the energy-momentum tensor. These have been found in Ref. \cite{correlation}. By $\rm{SU}(N)$ color symmetry, non vanishing form factors have the same number of particles and antiparticles. These are
\beq
\langle 0\!\!\!&\vert&\!\!\! T_{\mu\nu}(0)\vert A,\theta_1,b_1,a_1;\dots;A,\theta_M,b_M,a_M;P,\theta_{M+1},a_{M+1},b_{M+1};\dots;P,\theta_{2M},a_{2M},b_{2M}\rangle\nonumber\\
&=&[(p_1+\cdots+p_{2M})_\mu(p_1+\cdots+p_{2M})_\nu-\eta_{\mu\nu}(p_1+\cdots+p_{2M})^2]\nonumber\\
&&\times\frac{1}{N^{M-1}}\sum_{\sigma,\tau\in S_{M}}F^T_{\sigma\tau}(\theta_1,\dots,\theta_{2M})\prod_{j=1}^M\delta_{a_j a_{\sigma(j)+M}}\prod_{k=1}^M\delta_{b_k b_{\tau(k)+M}},\nonumber
\eeq
where $\sigma$ and $\tau$ are the permutations that take the numbers $1,\dots,M$ to $\sigma(1),\dots,\tau(M)$ and $\tau(1),\dots,\tau(M)$, respectively. At large $N$:
\beq
F^T_{\sigma\tau}(\theta_1,\dots,\theta_M)=\left\{\begin{array}{c}
\frac{(-2\pi^2)(4\pi)^{M-1}}{\prod_{j=1}^M(\theta_j-\theta_{\sigma(j)+M}+\pi i)\prod_{k=1}^M(\theta_k-\theta_{\tau(k)+M}+\pi i)},\,\,{\rm for} \,\sigma(j)\neq\tau(j),\,{\rm for\,all}\,j,\\
0,\,\,{\rm otherwise}\end{array}\right. .\label{tformfactor}
\eeq
We will be interested in the trace of the energy-momentum tensor operator, $\Theta=T_\mu^\mu$.

\section{Correlation Function and Asymptotic Freedom} 

Using the exact form factors, Eq.(\ref{formfactoransatz}) and (\ref{formfactorsolution}), an expression for the infinite-volume two-point correlation function of the renormalized field was written in Ref. \cite{renormalizedfield}. This correlation is found by summing over all the intermediate states:
\beq
\mathcal{W}(x)=\frac{1}{N}\sum_{a_0,b_0}\langle0\vert\Phi(x)_{b_0a_0}[\Phi_{b_0a_0}(0)]^*\vert0\rangle=\frac{1}{N}\sum_{a_0,b_0}\sum_{\Psi}e^{ix\cdot p_{\Psi}}\langle0\vert \Phi(0)_{b_0a_0}\vert\Psi\rangle\langle\Psi\vert[\Phi(0)_{b_0a_0}]^*\vert0\rangle,\label{intermediatestates}
\eeq
where $\Psi$ is any state with particles and antiparticles, and $p_\Psi$ is the sum of the momenta of the excitations of the state $\Psi$.

By directly introducing the exact form factors into (\ref{intermediatestates}), one finds
\beq
\mathcal{W}(x)=\sum_{M=1}^\infty\frac{1}{(M-1)!}\frac{1}{M!}\int\left(\prod_{j=1}^{2M-1}\frac{d\theta_j}{4\pi}\right)\sum_{\sigma\tau}\vert F_{\sigma\tau}(\theta_1,\dots,\theta_{2M-1})\vert^2\exp{\left(ix\cdot\sum_{j=1}^{2M-1} p_j \right)}+\mathcal{O}\left(\frac{1}{N}\right).\nonumber
\eeq
A drastic simplification comes from realizing that for a given value of $M$, all the pairs of permutations $\sigma,\tau$ give the same contribution to the correlation function. The final result of Ref.\cite{renormalizedfield} is (ignoring $\mathcal{O}(1/N)$ terms)
\beq
\mathcal{W}(x)=\frac{1}{4\pi}\sum_{l=0}^\infty\int_{-\infty}^{\infty} d\theta_1\dots\int_{-\infty}^{\infty} d\theta_{2l+1}\exp{\left(ix\cdot\sum_{j=1}^{2M-1} p_j \right)}\prod_{j=1}^{2l}\frac{1}{(\theta_j-\theta_{j+1})^2+\pi^2}.\label{correlationfinal}
\eeq

Recently, the short-distance behavior ($x\to0$) of the function (\ref{correlationfinal}) has been examined \cite{asymptoticfreedom}. This was done in Euclidean space by looking at $x^1=0$ and $x^0=iR$, such that $\exp ix\cdot p_j\to\exp-mR\cosh\theta_j$. The strategy is to realize that for small $R$, the function $\exp-mR\cosh\theta_j$ looks like a plateau, where it is approximately 1 for $-L<\theta_j<L$, and zero everywhere else, where $L=\ln\frac{1}{mR}$. This technique was first used to study the short-distance behavior of the Ising model \cite{shortdistising}.

For short distances the function (\ref{correlationfinal}) then becomes
\beq
\mathcal{W}(iR,0)=\frac{1}{4\pi}\sum_{l=0}^\infty\int_{-L}^{L} d\theta_1\dots\int_{-L}^{L} d\theta_{2l+1}\prod_{j=1}^{2l}\frac{1}{(\theta_j-\theta_{j+1})^2+\pi^2}.\label{shortdistancecorrelator}
\eeq
The function (\ref{shortdistancecorrelator}) was studied in \cite{asymptoticfreedom}, and it was shown that it diverges in a way that is consistent with what is expected from asymptotic freedom.

There is an alternate (and equivalent) way of examining the short distance behavior of the correlation function. The function $\mathcal{W}(x)$ diverges at the point $x=0$ when one performs the integrals over the rapidities. One solution is to simply introduce a cutoff in the rapidities, $\lambda$, ``by hand". One then finds
\beq
\mathcal{W}^\lambda(0)=\frac{1}{4\pi}\sum_{l=0}^\infty\int_{-\lambda}^{\lambda} d\theta_1\dots\int_{-\lambda}^{\lambda} d\theta_{2l+1}\prod_{j=1}^{2l}\frac{1}{(\theta_j-\theta_{j+1})^2+\pi^2}.\label{cutoffcorrelator}
\eeq
The function (\ref{cutoffcorrelator}) is exactly the same as (\ref{shortdistancecorrelator}), except we have replaced $L$ by $\lambda$. It is convenient to introduce the variables $u_j=\theta_j/\lambda$, such that
\beq
\mathcal{W}^\lambda(0)=\frac{\lambda}{4\pi}\sum_{l=0}^\infty\int_{-1}^1 du_1\dots\int_{-1}^{1} du_{2l+1}\prod_{j=1}^{2l}\frac{1}{\lambda[(u_j-u_{j+1})^2+\pi^2/\lambda^2]}.\label{uvariables}
\eeq

We define the function 
\beq
T(u_j,u_k)=\frac{1}{\lambda[(u_i-u_k)^2+\pi^2/\lambda^2]},\nonumber
\eeq
and an operator $\hat{T}$ and vector space $\left\{|u\rangle\right\}$, such that
\beq
\langle u_i|\hat{T}| u_k\rangle=T(u_i,u_k),\,\,\langle u_i|u_k\rangle=\delta(u_i-u_k),\,\,1=\int du_j|u_j\rangle\langle u_j|.\label{operator}
\eeq
In terms of the operator $\hat{T}$, the correlation function can be written as
\beq
\mathcal{W}^\lambda(0)=\frac{\lambda}{4\pi}\int du'du \langle u'|\frac{1}{1-\hat{T}^2}|u\rangle.\label{wightmanfield}
\eeq

The technique used in \cite{asymptoticfreedom} was to realize that the operator $\hat{T}$ can be written approximately in terms of the fractional Laplacian operator, $\Delta^{1/2}=\sqrt{-d^2/du^2}$. It is shown in \cite{asymptoticfreedom} that one can write
\beq
\hat{T}=e^{-\frac{\pi}{\lambda}H(\lambda)},\nonumber
\eeq
where $H(\lambda)$ is some operator that satisfies $H(\lambda)=\Delta^{1/2}+\mathcal{O}(1/\lambda)$. 
The fractional Laplacian satisfies the eigenvalue equation $\Delta^{1/2}\varphi_n(u)=\alpha_n\varphi_n(u)$, where $n=1,2,...$ and $0<\alpha_1<\alpha_2<\dots,$ with $\varphi_n(\pm 1)=0$.

The correlation function (\ref{uvariables}) can be written, for large $\lambda$, as
\beq
\mathcal{W}^\lambda(0)=\frac{\lambda}{4\pi}\int\sum_{n=1}^\infty\left\vert\int_{-1}^{1} du \varphi_n(u)\right\vert^2\frac{1}{1-e^{-2\pi\alpha_n/\lambda+\mathcal{O}(1/\lambda^2)}}=\frac{\lambda^2}{8\pi^2}\int\sum_{n=1}^\infty\left\vert\int_{-1}^{1} du \varphi_n(u)\right\vert^2\alpha_n^{-1}.\label{logsquared}
\eeq
The correlation in (\ref{logsquared}) is proportional to $\lambda^2$. This rapidity cutoff is related to a standard Euclidean momentum cutoff, $\Lambda$, by
\beq
m^2\sinh^2(\lambda)+m^2\cosh^2(\lambda)=\Lambda^2,\nonumber
\eeq
so that
\beq
 \lambda=\sinh^{-1}\left(\sqrt{\frac{\Lambda^2}{2m^2}-\frac{1}{2}}\right)=\ln\left(\sqrt{\frac{\Lambda^2}{2m^2}-\frac{1}{2}}+\sqrt{\frac{\Lambda^2}{2m^2}+\frac{1}{2}}\,\right)\approx \ln\left(\frac{\Lambda}{m}\right).\nonumber
\eeq
This dependence of the correlation function on the logarithm squared of the momentum cutoff is a confirmation of the asymptotic freedom of the model, and is predicted by perturbation theory \cite{perturbation}.

\section{The PCSM at finite volume}

We make the $x^1$ direction finite by imposing periodic boundary conditions $\Psi(x^1)=\Psi(x^1+V)$ on all wave functions, where $V$ is the one-dimensional volume. Placing the system in a finite volume discretizes the energy spectrum. The quantization condition depends on the exact S-matrix, and is found using the Bethe ansatz. This is, for an $n$-excitation state \cite{thermodynamical}
\beq
e^{ip_jV}\prod_{k\neq j}^{n}S(\theta_j-\theta_k)=\pm 1,\,\,\,\,j=1,2,\dots,n,\label{betheansatz}
\eeq
where we have suppressed all the color indices in the S-matrix for simplicity. The selection rules are $\pm=+$, for boson-like interactions, $S(0)=1$, and $\pm=-$ for the fermionic case, $S(0)=-1$. Equivalently, one can write
\beq
m V\sinh\theta_j+\sum_{k\neq j}^{n}\Delta(\theta_j-\theta_k)=2\pi l,\label{bethe}
\eeq
where $\Delta(\theta)=-i\ln S(\theta)$ and $l$ is an integer for bosonic interactions, and a half-integer in the fermionic case. Solving the equations (\ref{bethe}) one can find the discrete spectrum of rapidities, $\theta_j$.

At large $N$, a particle (antiparticle) can interact nontrivially with at most two other antiparticles (particles). The most nontrivial $n$-excitation state one can define is a chain of alternating particles and antiparticles, where the $j$-th particle (antiparticle) has one color index contraction with the $(j-1)$-st antiparticle (particle) and the $(j+1)$-st antiparticle (particle). Using the S-matrix, Eq. (\ref{sindex}), the Bethe quantization condition for the $j$-th particle at large $N$ is
\beq
m V\sinh\theta_j-i\ln\left(\frac{\theta_{j \,j-1}-\pi i}{\theta_{j\,j-1}+\pi i}\right)-i\ln\left(\frac{\theta_{j\,j+1}+\pi i}{\theta_{j\,j+1}-\pi i}\right)=2\pi l,\label{largenbethe}
\eeq
where we use the notation $\theta_{j k}=\theta_j-\theta_k$. There are only two (instead of $n-1$) terms coming from the S-matrix in Eq.(\ref{largenbethe}).

It is useful to find the spectrum in the thermodynamic limit, where $V,n\to\infty$, but their ratio is fixed. It can be shown \cite{thermodynamical} that in the thermodynamic limit, the quantization condition (\ref{bethe}) becomes
\beq
\epsilon(\theta)=mV\cosh(\theta)\mp\int\frac{d\theta^\prime}{2\pi}\varphi(\theta-\theta^\prime)\ln\left(1\pm e^{-\epsilon(\theta^\prime)}\right),\label{pseudoenergy}
\eeq
where $\varphi(\theta)=\frac{d}{d\theta}\Delta(\theta)$, and $\epsilon(\theta)$ is the so-called pseudo energy. The interpretation of the pseudo energy is that the ``dressed" energy of a particle of rapidity $\theta$ is given by $\epsilon(\theta)/V$ in the thermodynamic limit.

The thermodynamic limit of Eq. (\ref{largenbethe}) is trivial because the first term on the left-hand side dominates over the other two (because there are only two terms coming from the S-matrix, instead of $n-1$ terms as in the usual TBA). The TBA at large $N$ is therefore trivial. The pseudo energy of a particle of rapidity $\theta$ is simply given by $\epsilon(\theta)=mV\cosh\theta$ (because $\varphi(\theta-\theta^\prime)=0$ in (\ref{pseudoenergy})). This was noticed in Ref. \cite{kazakov}, where this result is used to declare (incorrectly, as we propose) the 't Hooft limit not physically interesting.

Once the pseudo energy is known, the ground-state energy, $E_0(V)$, and partition function,$Z(L,V)$ obtained from the TBA are \cite{thermodynamical}:
\beq
E_0(V)&=&\mp\int\frac{d\theta}{2\pi}m\cosh\theta\log(1\pm e^{-\epsilon(\theta)})\nonumber\\
Z(L,V)&=&\exp\left[-LE_0(V) \right],\label{partitiontba}
\eeq
where $L$ is the size of the $x^0$ direction. The partition function (\ref{partitiontba}) can be written equivalently as a sum over states:
\beq
Z(L,V)=\sum_{n=0}^{\infty}\frac{1}{n!}\int \frac{d\theta_1}{4\pi}\cdots\frac{d\theta_n}{4\pi}\langle\theta_1,\dots,\theta_n\vert\theta_1,\dots,\theta_n\rangle \prod_{i=1}^{n}e^{-\epsilon(\theta_i)},\label{partitionfunction}
\eeq
where the scalar products in (\ref{partitionfunction}), $\langle\theta_1,\dots,\theta_n\vert\theta_1,\dots,\theta_n\rangle,$ are those of a free bosonic or fermionic theory. The only effect of the interactions at finite volume is that the energies of the excitations are dressed. Since the pseudo energies are trivial at large $N$, the partition function derived from the TBA is that of an ideal gas. 
Despite this fact, it is easy to see why the expectation value of an operator $\mathcal{O}$, is not trivial. The expectation value can be formally written as
\beq
\langle \mathcal{O}\rangle^V=\frac{1}{Z(L,V)}\sum_{n=0}^{\infty}\frac{1}{n!}\int \frac{d\theta_1}{4\pi}\cdots\frac{d\theta_n}{4\pi}\langle\theta_1,\dots,\theta_n\vert\mathcal{O}\vert\theta_1,\dots,\theta_n\rangle \prod_{i=1}^{n}e^{-\epsilon(\theta_i)}.\label{formalexpectation}
\eeq
The expression (\ref{formalexpectation}) involves a sum over the form factors of the operator. As we have seen in the previous sections, the form factors are not trivial, even at large $N$. The expectation values are then different from those of a free theory. This expression has singularities that need to be regularized, before it can be used explicitly. In the rest of this paper we use the regularization scheme proposed by Leclair and Mussardo \cite{leclairmussardo}.
 We do not derive the Leclair-Mussardo (LM) formula here but simply quote it and use it.

It is easy to understand the failure of the TBA partition function if we don't suppress the color indices in the expression (\ref{partitionfunction}). The scalar products in (\ref{partitionfunction}) involve only the symmetric part of the S-matrix, while disregarding the effect of any nontrivial contractions of color indices between particles. At large $N$, the symmetric part of the S-matrix is trivial, so all the nontrivial information from the S-matrix is ignored. On the other hand, the form factors in (\ref{formalexpectation}) involve a sum over all the nontrivial color contractions. The main disparity is that the TBA partition function throws away the nontrivial color contractions, while the form factors do not.

One simple modification to the partition function (\ref{partitionfunction}) is to use the full S-matrix elements of the form (\ref{sindex}) to compute the scalar products, instead of using the scalar products from the free theory. The contributions from the antisymmetric part of the S-matrix are suppressed by higher powers of $1/N$. For example, the two-excitation contribution to the partition function involves the scalar product
\beq
\delta_{c_1}^{a_1}\delta_{c_2}^{a_2}\delta_{d_1}^{b_1}\delta_{d_2}^{b_2}S(\theta)_{a_1 b_1;b_2 a_2}^{d_2 c_2; c_1 d_2}=\left(N^2-\frac{2\pi i}{\pi i-\theta}\right)^2,\nonumber
\eeq
where the TBA partition function accounts only for the leading $N^4$ term. The subleading terms cannot be ignored if one is interested in computing correlation functions. This is because only the antisymmetric part of the S-matrix gives nontrivial contributions to the form factors. The terms that are suppressed in the partition function become the leading terms in the correlation functions.

 It is important to stress out the peculiarities of the large-$N$ limit that made the Bethe-ansatz calculation so simple. First, the large-$N$ S-matrix is diagonal, as was pointed out in \cite{correlation}. When two excitations scatter, they keep their identity (color charge quantum numbers), as the amplitude of an identity-changing process is suppressed by a factor of $1/N$. This is clearly seen by examining Eq. (\ref{sindex}). The only nontrivial scattering process is when two excitations have one or two color-index contractions. One can then forget about the color structure and reformulate the problem as a diagonal theory of particles that can interact with the S-matrices
\beq
S(\theta)=1, \frac{\theta\pm\pi i}{\theta\mp\pi i},\left(\frac{\theta\pm\pi i}{\theta\mp\pi i}\right)^2.\nonumber
\eeq
The second simplifying property of the large-$N$ limit is that, as we mentioned in Section II, there are no bound states. The binding energy of the bound states vanishes at large $N$, therefore the Bethe ansatz involves only elementary particles, with the scattering we just described.

The Bethe ansatz is significantly more complicated for arbitrary finite $N$. The particle spectrum consists of $N-1$ bound states with non-diagonal scattering. The product of S-matrices  in Eq.(\ref{betheansatz}) (commonly referred-to as the transfer matrix) has a complicated color structure and needs to be diagonalized. The eigenstates of the transfer matrix can be expressed in terms of the physical $N-1$ bound states, plus auxiliary ``magnon" particles that carry no energy or momentum. Furthermore, the Bethe equations for the auxiliary particles allow for additional magnon bound states, usually called ``strings".  One can then express the problem as a diagonal scattering theory, but including scattering with an infinite number of auxiliary string states. The necessary ingredients are all the S-matrices of physical bound states and strings.  A detailed derivation of the S-matrices and Bethe equations for the PCSM at general $N$ is found in \cite{leurentthesis}. 

It is necessary to understand what happens to the contribution from all these physical bound states and string states when we take the large-$N$ limit. The contribution from the string states to the partition function at finite $N$ is a result of their nontrivial scattering with the physical particles (The exact S-matrices can be found in Eq. (III.22) and (III.30) of \cite{leurentthesis}). From those expressions, it is  easy to see that at large $N$, these string-physical particle S-matrices become $1+\mathcal{O}(1/N)$. That is, the strings and physical particles stop interacting with each other at large $N$. There are then no contributions from auxiliary strings to the partition function

The disappearance of the contribution to the ground-state energy from the physical bound states is a bit more subtle. As we have mentioned, the S-matrix between two elementary excitations (with no color contractions), in the large-$N$ limit is $S(\theta)=1$, yielding the partition function of a free boson. However, for any finite $N$, the elementary particles satisfy fermionic selection rules, $S(0)=-1$. If one uses the finite-$N$ Bethe ansatz, one has to treat the particles as fermionic. If we later take the large-$N$ limit, we have to do so while using the fermionic rules. This process of taking the large-$N$ limit after computing the ground state energy, instead of before, means that the S-matrix of the elementary particles will actually be given by
\beq
S(\theta)=\left\{\begin{array}{c}
1,\,\,\,{\rm if}\,\,\theta\neq 0\\
-1,\,\,\,{\rm if}\,\,\theta=0\end{array}\right..\label{simplifiedmodel}
\eeq
The TBA for a theory with the S-matrix (\ref{simplifiedmodel}) has been examined in \cite{onleclairmussardo}. As is expected, and as is necessary for the consistency of our previous analysis, it is shown in \cite{onleclairmussardo} that the ground state energy and partition function of the model with S-matrix (\ref{simplifiedmodel}) is exactly the same as that of a free boson.

 As we have discussed before, the physical bound states dissolve into elementary particles in the planar limit. That is, the contribution to the ground state energy from an $r$-particle bound state becomes equivalent to the contribution of $r$ elementary particles. However, as this bound state is dissolved, the remaining elementary particles have the same rapidity. As we discussed, if we take the large-$N$ limit after computing the Bethe equations, the particles satisfy fermionic rules, and therefore we must enforce Pauli's exclusion principle. That is, a state with $r$-elementary particles with the same rapidity is not allowed, therefore there is no contribution to the ground state energy from the physical bound states.
 
 Treating the large-$N$ PCSM as a diagonal scattering theory from the beginning is a very useful shortcut. The more rigorous approach is to compute first the off-diagonal TBA and later take the large-$N$ limit, but this will yield the same free boson partition function.

 \section{The One-Point Function of the Energy-Momentum Tensor at Finite Volume}

In this section we evaluate the vacuum expectation value of the trace of the energy-momentum tensor at finite volume. We use the one-point function LM formula. This expectation value is interesting because it is usually easily calculated from the TBA. In field theories that are not matrix-valued, it has been shown that the results from the LM formula and those from the TBA agree \cite{leclairmussardo}. 

In our case, the TBA yields the expectation values of a free theory. Our position is that this is not the right value.  We believe the value of the LM formula is the correct one, as it uses the nontrivial form factors. We believe this discrepancy is simply a consequence of the field being a matrix.
Our approach then will be to find the expectation value of the energy-momentum tensor, assuming the validity of the LM formula, and then define a partition function such that it agrees with this value.

The LM one-point function for some operator $\mathcal{O}$ is
\beq
\langle \mathcal{O}\rangle^V=\sum_{n=0}^\infty\frac{1}{n!}\int_{\theta_1<\theta_2<\dots<\theta_n}\frac{d\theta_1}{2\pi}\cdots\frac{d\theta_n}{2\pi}\prod_{i=1}^n f_{-1}(\theta_i)\langle\theta_1,\dots,\theta_n\vert \mathcal{O}(0)\vert\theta_1,\dots,\theta_n\rangle_{\rm connected},\label{onepointlm}
\eeq
where $f_{\sigma_j}(\theta_j)=1/(1+e^{-\sigma_j\epsilon(\theta_j)})$, (the $\sigma_j=+1$ case will be relevant for the two point function). The connected form factor is defined as the finite part of the form factor after requiring that the rapidties of the incoming and outgoing states are equal. Any part of the form factors in (\ref{onepointlm}) that is divergent in this limit of the rapidities is discarded. This regularization is explained in more detail in \cite{leclairmussardo}.

We have used fermionic selection rules, $S(0)=-1$, in defining the functions $f_{\sigma_j}(\theta_j)$ because only the antisymmetric part of the S-matrix gives nontrivial contributions to the form factors. This antisymmetric part satisfies fermionic rules, while the symmetric part is bosonic. For a bosonic theory with $S(0)=1$, the corresponding functions in the LM formula would be $f_{\sigma_j}^{\rm bosonic}(\theta_j)=1/(1-e^{-\sigma_j\epsilon(\theta_j)})$.

We need the form factors of the operator $\Theta=T_\mu^\mu$ with the same number of excitations in the incoming and outgoing states. These can be obtained from (\ref{tformfactor}) by crossing symmetry.

The connected form factors of the energy-momentum tensor can be written neatly in terms of the S-matrix. We follow the calculation and language from Ref. \cite{leclairmussardo}. For a general scalar-valued field theory with S-matrix, $S(\theta)$, the connected form factors of the energy momentum tensor are \footnote{This equivalence holds only for our purposes, where all the rapidites are to be integrated, as in Eq. (\ref{onepointlm}). The factor of $n!$ accounts for the different permutations of the order of particle rapidities. All these permutations give the same contribution to the integral (\ref{onepointlm}).}
\beq
\langle\theta_1\dots \theta_n\vert \Theta\vert \theta_1\dots \theta_n\rangle_{\rm connected}=4\pi m^2 n!\,\varphi(\theta_{12})\varphi(\theta_{23})\cdots\varphi(\theta_{n-1,n})\cosh(\theta_{1n}),\label{energymomentumscalar}
\eeq
where $\varphi(\theta)=-i\frac{d\log S(\theta)}{d\theta}$.

The main difficulty when directly trying to apply Eq. (\ref{energymomentumscalar}) to our matrix-valued case is that the function $\varphi(\theta_{ij})$ is not the same for every pair of particles $i,j$. If the excitations $i$ and $j$ don't have any contracted color indices, the function $\varphi(\theta_{ij})$ vanishes. The only non-zero connected form factors are those where all the functions $\varphi(\theta_{j,j+1})$ are non-zero. We can build a state with alternating particles and antiparticles. The only color combinations that survive are those where the $j$-th particle has one color contraction with the $(j-1)$-st and the $(j+1)$-st antiparticles. The interaction between the $j$-th and the $(j+1)$-st excitations is given by the function
\beq
\varphi(\theta_{j, j+1})=-i\frac{d}{d\theta_{j,j+1}}\log\left(\frac{\theta_{j,j+1}+\pi i}{\theta_{j,j+1}-\pi i}\right)=-\frac{2\pi}{\theta_{j,j+1}^2+\pi^2}.\label{varphifunction}
\eeq

The non-vanishing connected form factors for our energy-momentum tensor are
\beq
&\frac{1}{N^2}\delta_{a_1^\prime a_1}\dots\delta_{a_n^\prime a_n}\delta_{b_1^\prime b_1}\dots\delta_{b_n^\prime b_n}\langle A,\theta_1,b_1^\prime,a_1^\prime;P,\theta_2,a_2^\prime,b_2^\prime;A,\theta_3,b_3^\prime,a_3^\prime;\dots\vert \Theta\vert A,\theta_1,b_1,a_1;P,\theta_2,a_2,b_2;\dots\rangle_{\rm connected}\nonumber\\
&=4\pi m^2n! \,\varphi(\theta_{12})\varphi(\theta_{23})\cdots\varphi(\theta_{n-1,n})\cosh(\theta_{1n})+\mathcal{O}\left(\frac{1}{N}\right).\label{pcsmconnected}
\eeq

The one-point function is found by substituting the form factors (\ref{pcsmconnected}) into the formula (\ref{onepointlm}). Our final result is
\beq
\frac{\langle \Theta\rangle^V}{N^2}=4\pi m^2\left(\sum_{n=1}^\infty\left[\prod_{i=1}^n\int\frac{d\theta_i}{4\pi}f_{-1}(\theta_i)\right]\left[\prod_{i=1}^{n-1}\frac{-2\pi}{\theta_{i,i+1}^2+\pi^2}\right]\cosh(\theta_{1n})\right)+\mathcal{O}\left(\frac{1}{N}\right).\label{finalonepoint}
\eeq

It is easy to see from Eq. (\ref{pcsmconnected}) and Eq. (\ref{finalonepoint}) why our results disagree with the trivial TBA. The difference between the connected form factors of a scalar theory (Eq. (\ref{energymomentumscalar})), and our matrix-valued case is that all the fundamental particles in a scalar theory interact with the same S-matrix. In the matrix-valued case, the S-matrix of two particles depends on how their colors are contracted. The Bethe equations of an $n$-excitation state involve the S-matrix of the $j$-th excitation with all other excitations. This is trivial in our case because excitations interact nontrivially with only two other excitations. The connected form factors for an $n$-excitation state, however, involve the S-matrix of each adjacent pair of particles $j$ and $j+1$. The $n$-particle state can be designed in such a way that all these two-particle S-matrices are nontrivial.

Our expectation value (\ref{finalonepoint}) can be used to define  a nontrivial partition function. The expectation value of the energy-momentum tensor is related to the finite-volume ground state energy, $E_0(V)$, by
\beq
\langle \Theta\rangle^V=\frac{2\pi}{V}\frac{d}{dV}\left[V E_0(V)\right].\label{groundstate}
\eeq
One can find the ground state energy in principle by solving the differential equation (\ref{groundstate}). The thermodynamic limit of the partition function is dominated by the ground state energy. We can then define the nontrivial thermodynamic limit of the partition function as
\beq
Z(L,V)=e^{-LE_0(V)},\nonumber
\eeq
where $L$ is the size of the $x^0$ direction.

 \section{Two-Point Correlation Function of the Renormalized Field at Finite Volume}

  In this Section we compute the two point correlation function of the renormalized field at finite volume.
For a local operator $\mathcal{O}(x)$ of an integrable theory, the LM two-point function is (again suppressing color indices)
\beq
\langle\Omega\vert\mathcal{O}(x)\mathcal{O}(0)\vert\Omega\rangle^V&=&\left(\langle\Omega\vert\mathcal{O}\vert\Omega\rangle^V\right)^2+\sum_{n=1}^\infty\frac{1}{n!}\sum_{\sigma_i=\pm1}\int\frac{d\theta_1}{4\pi}\dots\frac{d\theta_N}{4\pi}\left[\prod_{j=1}^nf_{\sigma_j}(\theta_j)\exp{\left(-\sigma_j\left(x^0\epsilon_j/V+ix^1k_j\right)\right)}\right]\nonumber\\
&&\times\vert\langle\Omega\vert\mathcal{O}(0)\vert\theta_1,\dots,\theta_n\rangle_{\sigma_1,\dots,\sigma_n}\vert^2,\label{leclairmussardo}
\eeq
where $k_j$ is the dressed finite volume momentum of the $i$-th particle (which at large $N$ is just the standard $k_j=m\sinh\theta_j$), and $\Omega$ is the dressed vacuum energy at finite volume. The first term in the right-hand side of (\ref{leclairmussardo}) is the squared expectation value of the operator at finite volume. The form factors used in (\ref{leclairmussardo}) are modified by the set of indices $\sigma_1,\dots,\sigma_n$. The meaning of this index is that if $\sigma_j=-1$, the $j$-th incoming particle (antiparticle) is crossed into an outgoing antiparticle (particle). All the excitations with $\sigma_j=1$ are in the incoming state.

We would like to point out that the validity of the LM two-point function has been questioned in References \cite{saleur} and \cite{fvform}. The main concern in Ref. \cite{saleur} is that the form factors used in the formula are those found at infinite volume, and they are not appropriate to find finite-volume correlation functions. The problem with using infinite-volume form factors is that the energies are dressed at finite volume. Thus when calculating finite-volume form factors one should use the appropriately dressed form factors. However, as we discussed before, at large $N$, the TBA pseudo energies of the PCSM are trivial. The pseudo energies from the TBA are those of a free theory, and 
``undressed", infinite-volume form factors seem appropriate. In this sense, our case is similar to free theories, where the LM formula is valid \cite{saleur}. A similar case is that of the thermal deformation of the Ising model. This is a theory of free massive fermions, and the two point functions were calculated in \cite{isingfinite}.

A different objection to the LM two-point formula is discussed in \cite{fvform}. The authors suggest that the series (\ref{leclairmussardo}) is not well defined for $n\geq3$. The form factors with both incoming and outgoing excitations have  poles at real values of the rapidities. Each rapidity has to be integrated over the real axis, and so, the integrals in (\ref{leclairmussardo}) are divergent, and not well defined in general.  Several regularization schemes for dealing with these divergences have been proposed \cite{esslerkonik},\cite{fvform}. In our model, however, we will see when crossing excitations to the outgoing state, the  poles are not pushed towards the real axis, and all our integrals are well defined. If two incoming excitations have a  pole at the rapidity difference $\theta=\pi i$, and one of these excitations is crossed into the outgoing state, the  pole is moved to $\theta=\pm 2\pi i$, instead of $\theta=0$. This is a consequence of the fact that our  poles are not periodic under $\theta\to\theta+2\pi i$, as were the usual  poles considered in \cite{fvform}. 

We do not have any further proof that the two-point LM formula is valid in our case, except that the usual objections against it do not apply. The main point we want to make is that the thermal correlation functions are not the trivial ones of a free theory. Even if the LM formula is not completely accurate, it is useful enough to show that the thermal correlators at large $N$ are not trivial.

We now find the general form factors needed for  (\ref{leclairmussardo}). Because of the global ${\rm SU}(N)\times {\rm SU}(N)$ symmetry of the PCSM, the non-vanishing form factors are
\beq
&&\langle A,\theta_{M+M^\prime+1},b_{M+M^\prime+1},a_{M+M^\prime+1};\dots;A,\theta_{M+M^\prime+k},b_{M+M^\prime+k},a_{M+M^\prime+k};P,\theta_{M+M^\prime+k+1},a_{M+M^\prime+k+1},b_{M+M^\prime+k+1};\nonumber\\
&&\dots;P,\theta_{M+M^\prime+k+k^\prime},a_{M+M^\prime+k+k^\prime},b_{M+M^\prime+k+k^\prime}\vert\Phi_{b_0a_0}(0)\vert A,\theta_1,b_1,a_1;\nonumber\\
&&\dots;A,\theta_M,b_M,a_M;P,\theta_{M+1},a_{M+1},b_{M+1};\dots;P,\theta_{M+M^\prime},a_{M+M^\prime},b_{M+M^\prime}\rangle,\label{crossedformfactor}
\eeq
with the condition $k+M^\prime-1=k^\prime+M$. We define permutations $\sigma,\tau\in S_{M+k^\prime}$ that take the set of numbers $\mathcal{A}=\{0,\dots,M,M+M^\prime+k+1,\dots,M+M^\prime+k+k^\prime\}$ to the set of numbers $\mathcal{B}=\{M+1,\dots,M+M^\prime+k\}$. With this notation we can express the form factor (\ref{crossedformfactor}) as
\beq
\sum_{\sigma,\tau\in S_{M+k^\prime}}\frac{1}{N^{M^\prime+k-\frac{1}{2}}}F_{\sigma\tau}(\theta_1,\dots,\theta_{M+M^\prime+k+k^\prime})\prod_{j=0}^{M}\delta_{a_j a_{\sigma(j)}}\delta_{b_jb_{\tau(j)}}\prod_{j=M+M^\prime+k+1}^{M+M^\prime+k+k^\prime}\delta_{a_j a_{\sigma(j)}}\delta_{b_jb_{\tau(j)}},\label{crossedformfactoransatz}
\eeq

We now introduce some further notation needed to write down a neat general expression for the function $F_{\sigma\tau}(\theta_1,\dots,\theta_{M+M^\prime+k+k^\prime})$. We define $\mathcal{A}_\sigma^1$, as the subset of $\mathcal{A}$, such that $\sigma(j)\in\{M+1,\dots, M+M^\prime\}$ for $j\in\{0,\dots,M\}$, %and $\sigma(j)\in\{M+M^\prime+1,\dots,M+M^\prime+k\}$ for $j\in\{M+M^\prime+k+1,\dots,M+M^\prime+k+k^\prime\}$,
 for all $j\in\mathcal{A}_\sigma^1$. Similarly $\mathcal{A}_\tau^1$ is defined such that $\tau(j)\in\{M+1,\dots, M+M^\prime\}$ for $j\in\{0,\dots,M\}$,% and $\tau(j)\in\{M+M^\prime+1,\dots,M+M^\prime+k\}$ for $j\in\{M+M^\prime+k+1,\dots,M+M^\prime+k+k^\prime\}$,
  for all $j\in\mathcal{A}_\tau^1$. We define $\mathcal{A}_{\sigma}^2\in\{0,\dots,M\}$ such that $\sigma(j)\in\{M+M^\prime+1,\dots,M+M^\prime+k\}$, for all $j\in\mathcal{A}_\sigma^2$, and $\mathcal{A}_\sigma^3\in\{M+M^\prime+k+1,\dots,M+M^\prime+k+k^\prime\}$, such that $\sigma(j)\in\{M+1,\dots,M+M^\prime\}$, for all $j\in\mathcal{A}_\sigma^3$. Finally we define $\mathcal{A}_\sigma^4\in \{M+M^\prime+k+1,\dots,M+M^\prime+k+k^\prime\}$, such that $\sigma(j)\in\{M+M^\prime+1,\dots,M+M^\prime+k\}$, for all $j\in\mathcal{A}_\sigma^4$. We similarly define $\mathcal{A}_\tau^2$, $\mathcal{A}_{\tau}^3$, and $\mathcal{A}_\tau^4$, in an analogous way. For a given pair of permutations $\sigma,\tau$, we define $n^1$ as the number of elements in the set $\mathcal{A}_\sigma^1$ plus the number of elements in the set $\mathcal{A}_{\tau}^1$. Similarly, we define $n^2$ as the number of elements in the sets $\mathcal{A}_{\sigma}^2$ and $\mathcal{A}_\tau^2$, $n^3$ is the number of elements in the sets $\mathcal{A}_\sigma^3$ and $\mathcal{A}_\tau^3$, and $n^4$ is the number of elements in the sets $\mathcal{A}_\sigma^4$ and $\mathcal{A}_\tau^4$. These numbers satisfy the condition $n^1+n^2+n^3+n^4=2(M+k^\prime)$ .
 
The general form factor, found from (\ref{formfactorsolution}) by using the S-matrix and crossing symmetry is given by
\beq
F_{\sigma\tau}(\theta_1,\dots,\theta_{M+M^\prime+k+k^\prime})&=&K_{\sigma\tau}\left[\prod_{j\in\mathcal{A}_{\sigma}^1,\mathcal{A}_\sigma^4}\left(\theta_j-\theta_{\sigma(j)}+\pi i\right)\prod_{j\in\mathcal{A}_\tau^1,\mathcal{A}_\tau^4}\left(\theta_j-\theta_{\tau(j)}+\pi i\right)\right.\nonumber\\
&&\times\left.\prod_{j\in\mathcal{A}_\sigma^2}\left(\theta_j-\theta_{\sigma(j)}+2\pi i\right)\prod_{j\in\mathcal{A}_\tau^2}\left(\theta_j-\theta_{\tau(j)}+2\pi i\right)\right.\nonumber\\
&&\times\left.\prod_{j\in\mathcal{A}_\sigma^3}\left(\theta_j-\theta_{\sigma(j)}-2\pi i\right)\prod_{j\in\mathcal{A}_\tau^3}\left(\theta_j-\theta_{\tau(j)}-2\pi i\right)\right]^{-1},\label{crossedformfactorsolution}
\eeq
where
\beq
K_{\sigma\tau}=\left\{\begin{array}{c}
(-4\pi)^{M+k^\prime}, \,\,\,\,\sigma(j)\neq\tau(j),\,{\rm for\,all\,}j,\\
0,\,\,\,\,{\rm otherwise}\end{array}\right. .\nonumber
\eeq

We now substitute the exact form factors into the LM formula (\ref{leclairmussardo}). After some tedious but straight forward calculation, we find that the finite-volume correlation function , for the operator $\mathcal{O}=\Phi/\sqrt{N}$, is 
\beq
\mathcal{W}(x)^V&=&\frac{1}{4\pi}\sum_{l=0}^{\infty}\sum_{n^1=0}^{2l}\,\sum_{n^2=0}^{2l-n^1}\sum_{n^4=0}^{2l-n^1-n^2}\int_{-\infty}^{\infty}d\theta_{1}\dots \int_{-\infty}^\infty d\theta_{2l+1}\left[f_1(\theta_1)f_1(\theta_{2l+1})\right]^{\frac{1}{2}}\nonumber\\
&&\times\prod_{j=1}^{n^1}\frac{\left[f_{1}(\theta_j)f_{1}(\theta_{j+1})\exp\left\{-[t(\epsilon_j+\epsilon_{j+1})/V+ix(k_j+k_{j+1})]\right\}\right]^{\frac{1}{2}}}{(\theta_j-\theta_{j+1})^2+\pi^2}\nonumber\\
&&\times\prod_{j=n^1+1}^{n^1+n^2}\frac{\left[f_{1}(\theta_j)f_{-1}(\theta_{j+1})\exp\left\{-[t(\epsilon_j-\epsilon_{j+1})/V+ix(k_j-k_{j+1})]\right\}\right]^{\frac{1}{2}}}{(\theta_j-\theta_{j+1})^2+4\pi^2}\nonumber\\
&&\times\prod_{j=n^1+n^2+1}^{n^1+n^2+n^4}\frac{\left[f_{-1}(\theta_j)f_{-1}(\theta_{j+1})\exp\left\{[t(\epsilon_j+\epsilon_{j+1})/V+ix(k_j+k_{j+1})]\right\}\right]^{\frac{1}{2}}}{(\theta_j-\theta_{j+1})^2+\pi^2}
\nonumber\\
&&\times\prod_{n^1+n^2+n^4+1}^{2l}\frac{\left[f_{-1}(\theta_j)f_{1}(\theta_{j+1})\exp\left\{-[t(-\epsilon_j+\epsilon_{j+1})/V+ix(-k_j+k_{j+1})]\right\}\right]^{\frac{1}{2}}}{(\theta_j-\theta_{j+1})^2+4\pi^2}
\nonumber\\
&&+\frac{1}{4\pi}\sum_{l=0}^{\infty}\sum_{n^1=0}^{2l}\,\sum_{n^2=0}^{2l-n^1}\sum_{n^3=0}^{2l-n^1-n^2}\int_{-\infty}^{\infty}d\theta_{1}\dots \int_{-\infty}^\infty d\theta_{2l+1}\left[f_{-1}(\theta_1)f_{-1}(\theta_{2l+1})\right]^{\frac{1}{2}}\nonumber\\
&&\times\prod_{j=1}^{n^3}\frac{\left[f_{-1}(\theta_j)f_{1}(\theta_{j+1})\exp\left\{-[t(-\epsilon_j+\epsilon_{j+1})/V+ix(-k_j+k_{j+1})]\right\}\right]^{\frac{1}{2}}}{(\theta_j-\theta_{j+1})^2+4\pi^2}\nonumber\\
&&\times\prod_{j=n^3+1}^{n^1+n^3}\frac{\left[f_{1}(\theta_j)f_{1}(\theta_{j+1})\exp\left\{-[t(\epsilon_j+\epsilon_{j+1})/V+ix(k_j+k_{j+1})]\right\}\right]^{\frac{1}{2}}}{(\theta_j-\theta_{j+1})^2+\pi^2}\nonumber\\
&&\times\prod_{j=n^1+n^3+1}^{n^1+n^2+n^3}\frac{\left[f_{1}(\theta_j)f_{-1}(\theta_{j+1})\exp\left\{-[t(\epsilon_j-\epsilon_{j+1})/V+ix(k_j-k_{j+1})]\right\}\right]^{\frac{1}{2}}}{(\theta_j-\theta_{j+1})^2+4\pi^2}\nonumber\\
&&\times\prod_{j=n^1+n^2+n^3+1}^{2l}\frac{\left[f_{-1}(\theta_j)f_{-1}(\theta_{j+1})\exp\left\{[t(\epsilon_j+\epsilon_{j+1})/V+ix(k_j+k_{j+1})]\right\}\right]^{\frac{1}{2}}}{(\theta_j-\theta_{j+1})^2+\pi^2}+\mathcal{O}\left(\frac{1}{N}\right)
.\label{finitevolumecorrelator}
\eeq

We now want to study how this correlation function diverges at $x=0$. As we did in the previous section, we will take $x=0$ and introduce a rapidity cutoff to regularize any divergence. The function (\ref{finitevolumecorrelator}) becomes
\beq
&&\mathcal{W}^\lambda(0)^V=\frac{1}{4\pi}\sum_{l=0}^{\infty}\sum_{n^1=0}^{2l}\,\sum_{n^2=0}^{2l-n^1}\sum_{n^4=0}^{2l-n^1-n^2}\int_{-\lambda}^{\lambda}d\theta_{1}\dots \int_{-\lambda}^\lambda d\theta_{2l+1}\left[f_1(\theta_1)f_1(\theta_{2l+1})\right]^{\frac{1}{2}}\nonumber\\
&&\times\prod_{j=1}^{n^1}\frac{\left[f_{1}(\theta_j)f_{1}(\theta_{j+1})\right]^{\frac{1}{2}}}{(\theta_j-\theta_{j+1})^2+\pi^2}\prod_{j=n^1+1}^{n^1+n^2}\frac{\left[f_{1}(\theta_j)f_{-1}(\theta_{j+1})\right]^{\frac{1}{2}}}{(\theta_j-\theta_{j+1})^2+4\pi^2}\,\prod_{j=n^1+n^2+1}^{n^1+n^2+n^4}\frac{\left[f_{-1}(\theta_j)f_{-1}(\theta_{j+1})\right]^{\frac{1}{2}}}{(\theta_j-\theta_{j+1})^2+\pi^2}\prod_{j=n^1+n^2+n^4+1}^{2l}\frac{[f_{-1}(\theta_j)f_{1}(\theta_{j+1})]^{\frac{1}{2}}}{(\theta_j-\theta_{j+1})^2+4\pi^2}\nonumber\\
&&+\frac{1}{4\pi}\sum_{l=0}^{\infty}\sum_{n^1=0}^{2l}\,\sum_{n^2=0}^{2l-n^1}\sum_{n^3=0}^{2l-n^1-n^2}\int_{-\lambda}^{\lambda}d\theta_{1}\dots \int_{-\lambda}^\lambda d\theta_{2l+1}\left[f_{-1}(\theta_1)f_{-1}(\theta_{2l+1})\right]^{\frac{1}{2}}\nonumber\\
&&\times\prod_{j=1}^{n^3}\frac{\left[f_{-1}(\theta_j)f_{1}(\theta_{j+1})\right]^{\frac{1}{2}}}{(\theta_j-\theta_{j+1})^2+4\pi^2}\prod_{j=n^3+1}^{n^1+n^3}\frac{\left[f_{1}(\theta_j)f_{1}(\theta_{j+1})\right]^{\frac{1}{2}}}{(\theta_j-\theta_{j+1})^2+\pi^2}\,\prod_{j=n^1+n^3+1}^{n^1+n^2+n^3}\frac{\left[f_{1}(\theta_j)f_{-1}(\theta_{j+1})\right]^{\frac{1}{2}}}{(\theta_j-\theta_{j+1})^2+4\pi^2}\prod_{j=n^1+n^2+n^3+1}^{2l}\frac{[f_{-1}(\theta_j)f_{-1}(\theta_{j+1})]^{\frac{1}{2}}}{(\theta_j-\theta_{j+1})^2+\pi^2}
.\nonumber\\
&&\label{regularizedfinitevolumecorrelator}
\eeq

In the following section we evaluate the expression (\ref{regularizedfinitevolumecorrelator}) taking two very different limits. First we examine (\ref{regularizedfinitevolumecorrelator})   at very large volume, $V\to\infty$, and recover the previous results from Section III. We then examine the opposite limit of very small volume, $V\to0$.

\section{The p-regime vs. the $\epsilon$-regime}

In this section we examine the function (\ref{regularizedfinitevolumecorrelator}) for both very large and very small volumes. There is only one length scale in the PCSM, namely $m$. By large volume, it is meant, that $V>>1/m$. In the finite-volume-QCD literature \cite{finitevolume}, this is commonly called the p-regime. By small volume, it is meant that $V<<1/m$, which is commonly called the $\epsilon$-regime. If instead the direction $x^0$ is made finite, the p-regime and the $\epsilon$-regime correspond to the low-temperature, and the high-temperature limit, respectively.

The volume dependence of the expression (\ref{regularizedfinitevolumecorrelator}) is included only in the functions 
\beq
f_{\sigma}(\theta)=\left\{\begin{array}{c}
\frac{1}{1+e^{-Vm\cosh\theta}},\,\,\,\,\sigma=1,\\
\,\\
\frac{e^{-Vm\cosh\theta}}{1+e^{-Vm\cosh\theta}},\,\,\,\,\sigma=-1\end{array}\right.\,\,.\nonumber
\eeq
Taking the large-volume limit, these become
\beq
\lim_{V\to\infty}f_{\sigma}(\theta)=\left\{\begin{array}{c}
1,\,\,\,\,\sigma=1,\\
\,\\
0,\,\,\,\,\sigma=-1\end{array}\right.\,\,.\nonumber
\eeq
Then at large volume,
equation (\ref{regularizedfinitevolumecorrelator}) becomes (\ref{wightmanfield}) and we simply recover the results we calculated at infinite volume. We consider this limit as a trivial consistency check of the LM formula.

Now we examine (\ref{regularizedfinitevolumecorrelator}) for $V<<1/m$. The argument we will use is similar to the one used to find the expression (\ref{shortdistancecorrelator}). We use the fact that for very small $V$, the function $e^{-Vm\cosh\theta}$ becomes approximately a plateau, with value 1 for $-\mathcal{L}<\theta<\mathcal{L}$, and 0 elsewhere, where $\mathcal{L}=\ln\frac{1}{mV}$. In this limit, then
\beq
f_1(\theta)=\left\{\begin{array}{c}
\frac{1}{2},\,\,\,\,-\mathcal{L}<\theta<\mathcal{L},\\
\,\\
1,\,\,\,\,{\rm otherwise},\end{array}\right.\,\,\,,\,\,\,\,\,\,\,\,\,\,\,\,\,\,\,
f_{-1}(\theta)=\left\{\begin{array}{c}
\frac{1}{2},\,\,\,\,-\mathcal{L}<\theta<\mathcal{L},\\
\,\\
0,\,\,\,\,{\rm otherwise}.\end{array}\right.\,\,\,.\nonumber
\eeq

Defining the new operators
\beq
\langle\theta\vert\hat{t}^1\vert\theta^\prime\rangle=\frac{1}{2(\theta-\theta^\prime)^2+2\pi^2},\,\,\,\,\langle\theta\vert\hat{t}^2\vert\theta^\prime\rangle=\frac{1}{2(\theta-\theta^\prime)^2+8\pi^2},\,\,\,\,\langle\theta\vert\hat{t}\vert\theta^\prime\rangle=\frac{1}{(\theta-\theta^\prime)^2+\pi^2},\nonumber
\eeq
the expression (\ref{regularizedfinitevolumecorrelator}) becomes
\beq
\mathcal{W}^\lambda(0)^V&=&\frac{1}{4\pi}\int_{-\mathcal{L}}^{\mathcal{L}} d\theta'd\theta \langle\theta'\vert\left(\frac{1}{1-(\hat{t}^1)^2}\right)^2\left(\frac{1}{1-(\hat{t}^2)^2}\right)^2\vert\theta\rangle\nonumber\\
&&+\frac{1}{4\pi}\int_{-\lambda}^{-\mathcal{L}}d\theta'd\theta \langle\theta'|\frac{1}{1-(\hat{t}\,)^2}|\theta\rangle+\frac{1}{4\pi}\int_{\mathcal{L}}^{\lambda}d\theta'd\theta \langle\theta'|\frac{1}{1-(\hat{t}\,)^2}|\theta\rangle\nonumber\\
&&\nonumber\\
&=&\frac{1}{4\pi}\int_{-\mathcal{L}}^{\mathcal{L}} d\theta'd\theta \langle\theta'\vert\left(\frac{1}{1-(\hat{t}^1)^2}\right)^2\left(\frac{1}{1-(\hat{t}^2)^2}\right)^2-\frac{1}{1-\hat{t}^2}\vert\theta\rangle\nonumber\\
&&+\mathcal{W}^\lambda(0).
\label{smallvolumecorrelator}
\eeq
The first term in the right-hand side of (\ref{smallvolumecorrelator}) has no dependence on $\lambda$.

We now express the result (\ref{smallvolumecorrelator}) in terms of the eigenvalues and eigenfunctions of the fractional Laplacian operator. We switch to new variables $u_j=\theta_j/\mathcal{L}$. For very large $\mathcal{L}$ (deep in the $\epsilon$-regime), Eq. (\ref{smallvolumecorrelator}) can be written as
\beq
\mathcal{W}^\lambda(0)^V-\mathcal{W}^\lambda(0)=\frac{\mathcal{L}}{4\pi}\int_{-1}^1du^\prime du\,\langle u^\prime\vert \left(\frac{1}{1-\frac{\hat{T}^2}{2}}\right)^4-\frac{1}{1-\hat{T}^2}\vert u\rangle=-\frac{\mathcal{L}^2}{8\pi^2}\int\sum_{n=1}^\infty\left\vert\int_{-1}^{1} du \varphi_n(u)\right\vert^2\alpha_n^{-1}+\mathcal{O}\left(\mathcal{L}\right)<0,\nonumber
\eeq
where the operator, $\hat{T}$, was defined in Eq. (\ref{operator}). It is important to notice that in the $\epsilon$-regime, the difference $\delta W^\lambda(0)=\mathcal{W}^\lambda(0)^V-\mathcal{W}^\lambda(0)$ is always negative. This means that at very small volumes (or very high temperatures) the correlation function becomes less and less divergent at $x=0$.

\section{Conclusions}

We have computed the thermal expectation value of the trace of the energy-momentum tensor of the PCSM at large $N$, using the Leclair-Mussardo formula.  This value is usually easy to determine from the TBA, which in our model is that of a free theory. The value we obtain from the LM formula is not trivial, and does not agree with what is expected from the TBA. This implies that there is a nontrivial ground state energy. The discrepancy arises from the fact that the two-particle S-matrix is nontrivial if the particles share a color contraction. The TBA only takes into account the symmetric part of the S-matrix, which is trivial at large $N$.

We have also calculated the two-point correlation function of the renormalized field operator  in a finite volume. This calculation was done using the LM formula for two-point functions. The validity of this formula has been questioned before. However, we argued that the usual objections do not apply in our particular case. In our case, all the integrals in the LM formula are well defined, and pseudo energies from the TBA are those of a free theory.  For very large volumes (in the p-regime), we recover the standard, infinite-volume two-point function from \cite{renormalizedfield},\cite{asymptoticfreedom}, which diverges logarithmically at short distances. For very small volumes (in the $\epsilon$-regime), we saw that this divergence gets softened as we reduce the volume size.

The ground state energy (and some excited states) of the finite-volume PCSM for some small values of $N$ have been calculated before in Ref. \cite{kazakovleurent} by solving the associated Hirota equation.  The large-N limit extrapolation of these results is not yet well understood. These results, however, only use the symmetric part of the S-matrix, so it is doubtful that our nontrivial results can be reproduced by simply extrapolating to large $N$. The authors of Ref. \cite{kazakovleurent} propose their results can be used to study the alternate large-$N$ limit of Ref. \cite{kazakov}. 

One might hope to obtain a nontrivial partition function and expectation values from the TBA by working at general finite $N$, and taking the large-$N$ limit only at the end of the calculation. This exercise is much harder than our case, since the Bethe equations are highly nontrivial. However, we argued that this process should yield the same free boson partition function, as long as we take the large-$N$ limit described in this paper, and not the one from  \cite{kazakov}

Our interpretation of our results is that the TBA partition function, starting from the large-$N$ limit of the S-matrix, is not enough to describe all the thermodynamics of a matrix-valued theory. There are contributions to vacuum expectation values of operators that arise from the matrix structure of the fields, which contains information ignored by this partition function.

\begin{acknowledgements}
I would like to thank Peter Orland for  many helpful discussions, particularly about the Bethe ansatz at large-$N$, and the asymptotic freedom of the PCSM. I also thank Ra\'ul Brice\~no for reading and commenting on an early version of this manuscript,  and giving me some introduction to the ideas of the finite-volume QCD literature. Finally I thank Giuseppe Mussardo, Gabor Takacs and Fabio Franchini for some comments and discussions of the final results.
This work has been supported by the ERC, under grant number 279391 EDEQS.

\end{acknowledgements}


\begin{thebibliography}{xx}

\bibitem{wiegmann} A.M. Polyakov and P.B. Wiegmann, Phys. Lett. {\bf 131 B} (1983) 121;  
E. Abadalla, M.C.B. Abadalla and M. Lima-Santos, Phys. Lett. {\bf 140 B} (1984) 71; P.B. 
Wiegmann, Phys. Lett. {\bf 141 B} (1984) 217; Phys. Lett. {\bf 142 B} (1984) 173.
\bibitem{unsal} A. Cherman, D. Dorigoni, G. V. Dunne and M. Unsal; Phys.Rev.Lett. {\bf 112} (2014) 021601. A. Cherman, D. Dorigoni and M. Unsal; arXiv:1403.1277 (2014).
\bibitem{renormalizedfield}P. Orland, Phys. Rev. {\bf D 84} (2011) 105005; Phys. Rev. {\bf D 86} (2012) 045023.

\bibitem{multiparticle}A. Cort\'es Cubero, Phys. Rev. {\bf D 86} (2012) 025025.
\bibitem{correlation} A. Cort\'es Cubero and P. Orland, Phys. Rev. {\bf D 88} (2013) 025044. There is a error in the expressions of the form factors in this reference, the correct version is found in: A. Cort\'es Cubero; arXiv:1409.8341, PhD Thesis, Graduate School and University Center of the City University of New York (2014).
\bibitem{asymptoticfreedom} P. Orland; Phys. Rev. {\bf D 90}, (2014) 125038.
\bibitem{leclairmussardo} A. LeClair and G. Mussardo, Nucl. Phys. {\bf B552}(1999) 624-642.
\bibitem{saleur} H. Saleur; Nucl. Phys. {\bf B567} (200) 602-610. O. A. Castro-Alvaredo and A. Fring; Nucl. Phys. {\bf B636} (2002) 611-631.
\bibitem{fvform}B. Pozsgay and G. Takacs; J.Stat.Mech. {\bf 1011} (2010) P11012.



\bibitem{bootstrap} F.A. Smirnov, 
{\bf Form Factors in Completely Integrable Models of Quantum Field Theory,} 
Advanced Series in Mathematical Physics, Vol. {\bf14}, World Scientific (1992). H. Babujian, A. Foerster and M. Karowski, Journ. Phys. {\bf A 41} (2008) 275202; Nucl. Phys. {\bf B 825} (2010) 396.
\bibitem{kazakov} V. Fateev, V. Kazakov and P. Wiegmann, Nucl. Phys. {\bf B 424} (1994) 505; V. Fateev, V. Kazakov and P. Wiegmann, Phys. Rev. Lett {\bf 73} (1994) 1750.
\bibitem{shortdistising} J.L. Cardy and G. Mussardo, Nucl. Phys.{\bf B 340} (1990) 387; V.P. Yurov and Al.B. Zamolodchikov, Int. J. Mod. Phys. {\bf A 6} (1991) 3419.
\bibitem{perturbation} A.M. Polyakov, {\bf Gauge Fields and Strings}, Sections 2.1 and 8.1, Harwood Academic Pulishers, Chur (1987). There is an
error in Eq. (8.36) of this book; The correct result
is found in P. Rossi and E. Vicari, Phys. Rev. {\bf D 49} (1994)
6072; Eq. (168), and in P. Rossi, M. Campostrini,
and E. Vicari, Phys. Rep. {\bf 302},  (1998) 143, Eqs. (7.5) and
(7.6).
\bibitem{thermodynamical} C.N. Yang and C.P. Yang, J. Math. Phys. {\bf 10} (1969),1115. Al.B. Zamolodchikov; Nucl. Phys.  {\bf B342} (1990), 695.

\bibitem{leurentthesis}S. Leurent; arXiv:1206.4061, PhD Thesis, L'Universit\'e Perre et Marie Curie (2012).

\bibitem{onleclairmussardo} G. Mussardo; J.Phys. {\bf A34} (2001) 7399-7410.


\bibitem{isingfinite}A. LeClair, F. Lesage, S. Sachdev and H. Saleur; Nucl. Phys. {\bf B 482} [FS] (1996) 579.
\bibitem{esslerkonik} F. H. L. Essler and R. M. Konik; J.Stat.Mech. {\bf 0909} (2009) P09018. 

\bibitem{finitevolume} S. Aoki and H. Fukaya; Phys.Rev. {\bf D84} (2011) 014501. A review of both the p-regime and the $\epsilon$-regime is given, and a proposal to interpolate between the two. Further references on the techniques used on each regime can be found here.
\bibitem{kazakovleurent} V. Kazakov and S. Leurent, arXiv: 1007.1770 (2010).























\end{thebibliography}
\end{document}